 \documentclass{elsart}
\usepackage{graphicx}

\def\prl {Phys. Rev. Lett. }  
\def\pre {Phys. Rev. E }

\def \be {\begin{equation}}
\def \ee {\end{equation}}
\def \d  {\delta}

\begin{document}
\begin{frontmatter}
\title{Critical Behavior of Sandpile Models with Sticky Grains}
\author{ P. K. Mohanty}
\address{TCMP Division, Saha Institute of Nuclear Physics, 1/AF Bidhannagar, Kolkata- 700064, 
INDIA}
\author{Deepak Dhar}
\address{Tata Institute of Fundamental Research, Homi Bhabha Road, 
Mumbai-400 005, INDIA}

\maketitle 
\begin{abstract} 

We revisit the question whether the critical behavior of sandpile models
with sticky grains is in the directed percolation universality class.  
Our earlier theoretical arguments in favor, supported by evidence from
numerical simulations [ Phys. Rev. Lett., {\bf 89} (2002) 104303], have
been disputed by Bonachela et al. [Phys.  Rev. E {\bf 74} (2004)  050102] for
sandpiles with no preferred direction. We discuss possible reasons for the
discrepancy. Our new results of longer simulations of the one-dimensional
undirected model fully support our earlier conclusions.

\end{abstract}
\end{frontmatter}

  After the pioneering work of Bak, Tang and Wiesenfeld\cite{btw} in 1987,
sandpile models have been studied extensively in the past two decades,
both as paradigms of self-organized critical systems in general\cite{soc},
and also as models of real granular matter \cite{granular}.  Many
different types of sandpile models with different toppling rules have been
studied \cite{kadanoff} : deterministic and stochastic, with or without
preferred direction, different instability criteria \cite{manna1}, or
particle distribution rules \cite{maslov}, with fixed energy
\cite{vespignani} etc.. Most of these models could only be studied
numerically, and for a while it seemed that each new variation studied
belonged to a new universality class of critical behavior.  
Though not complete, a broad picture
of the different universality classes of self-organized critical
behavior has emerged in recent years \cite{biham,brazil}.

 In an earlier paper \cite{mohanty_dhar}, we have argued that the generic
behavior of sandpile models is in the universality class of directed
percolation (DP), and models with deterministic toppling rules like the
original BTW model, and models with stochastic toppling rules like the
Manna models, are unstable to a perturbation of introduction of
``stickiness'' in the toppling rules, and under renormalization, the flows
are directed towards the DP-fixed point.  These arguments are reasonable,
but not rigorous, and we presented a detailed study of a specific model,
where some of the steps in the arguments could be shown to be valid, and
we used detailed Monte Carlo simulations to check our conclusions.

 Some of the conclusions of this paper have recently been disputed by
Bonachela et al. \cite{bonachela}.  These authors contend that while 
the directed sandpiles with sticky grains show DP behavior in the SOC
limit, our arguments do not apply to {\it undirected} sandpiles, where, 
even with stickiness, the
critical behavior continues be the same as that of the Manna model, i. e., 
in the universality class of the directed percolation with a  conservation 
law (hereafter referred to as the Manna/C-DP universality class).

In this paper, we will try discussing these conflicting claims, and also
present some data from more recent extensive simulations, which supports
our original conclusions.  We start by defining the model precisely, and
then summarize the arguments of \cite{mohanty_dhar}. We then discuss the
simulations of \cite{bonachela}, and finally present the results of our
new more extensive simulations.

 First the precise definition of the model.  We consider the directed
model on a $(1+1)$-dimensional square lattice, for definiteness.
Generalizations to higher dimensions are straight forward.  The sites on
an $L \times M$ torus are labelled by euclidean coordinates $(i,j)$ with
$(i+j)$ even and $j$ increasing downward. At each site $(i,j)$, there is
a non-negative integer $h_{i,j}$ to be called the height of the pile at
that site. Initially all $h_{i,j}$ are zero.  The system is driven by
choosing a site at random and increasing the height at that site by one.

The `stickiness' of the grains is characterized by a parameter $p$, and
its role in the dynamics of sandpiles is defined as follows:  A site is
said to become unstable at time $t$, if at least one particle is added to
it at time $t$, and its height becomes greater than $1$.  A site $(i,j)$
made unstable at time $t$ relaxes at the time $(t+1)$ stochastically: With
probability $( 1-p)$, it becomes stable {\it without losing any grains},
and the added particle(s) sticks to the existing grains.  Otherwise (with
probability $p$), the site topples, and the height at the site decreases
by two, and the site {\it becomes stable}. We introduce bulk dissipation:
at each toppling, with probability $\delta$ both grains from the toppling
are lost, other wise (with probability $1-\delta$), the two grains are
transferred to the two downward neighbors $(i \pm 1,j+1)$.

 We relax all the unstable sites by parallel dynamics.  An unstable site
is relaxed in one step, independent of whether it received one or more
grains at the previous time step. Once a site has relaxed, it remains
stable until perturbed again by new grains coming to the site. This
relaxation process is repeated until all sites become stable, and then a
new grain is added.

 The model is specified by two parameters $p$ and $\delta$. For this 
model, the following results can be proved \cite{mohanty_dhar}:\\
\begin{itemize}
\item[i)] Depending on the values of $p$ and $\delta$, two different behaviors 
are possible. There is a threshold  $p^*(\delta)$, such that for $ p> 
p^*(\delta) $, there is a steady state in the system, but for $p < 
p^*(\delta)$, no steady state is possible, and the mean height of the pile 
grows linearly with time [Fig. 1].\\
\item[ii)] The boundary line $p = p^*(\delta)$ is exactly given in terms of the 
function $\bar{n}_{DP}(\lambda)$, which gives the 
mean number of infected  and boundary sites in a cluster in a directed 
site percolation process 
on the same lattice with infection probability $\lambda$. In particular, 
this boundary line meets the $\delta = 0$ line at $p =p_c$, where $p_c$ is 
the exact directed  site percolation threshold for the square lattice.  \\
\item[iii)] At the boundary line $p = p^*(\delta)$, the distribution of sizes of 
avalanches is exactly the same as in the DP process, with a infection 
probability $\lambda = p(1 - \delta)$.\\
\item[iv)] In the regime where the steady state exists, mean cluster size is 
finite for non-zero $\delta$, and we get SOC behavior only for $ p_c \leq 
p \leq 1$, and $\delta \rightarrow 0$.\\
\item[v)] The model can be solved exactly along three of the boundary lines,  
$p=1$, $\d =1$ and $p=0$.
\end{itemize}

We find that correlations between heights at different sites in the steady
state are quite weak, and the steady state is almost a product measure
state. 
We argued that if these correlations are {\it irrelevant}, one can
study the avalanche process in  a background where these correlations
are {\it absent}. Then the avalanche process becomes a Domanay- Kinzel
process with two parameters $(p_1, p_2)$,
where $p_2=p$ and $p_1$ is determined in terms of the concentration of 
sites with height zero in the steady state. If the
correlations in height are indeed irrelevant, the avalanche distribution
function for all $p$ along the SOC line in Fig. 1 would show DP exponents,
except the end point $p=1, \delta =0$, where the model has deterministic
toppling rules.

\begin{figure}
\begin{center}
\includegraphics[width=4 in]{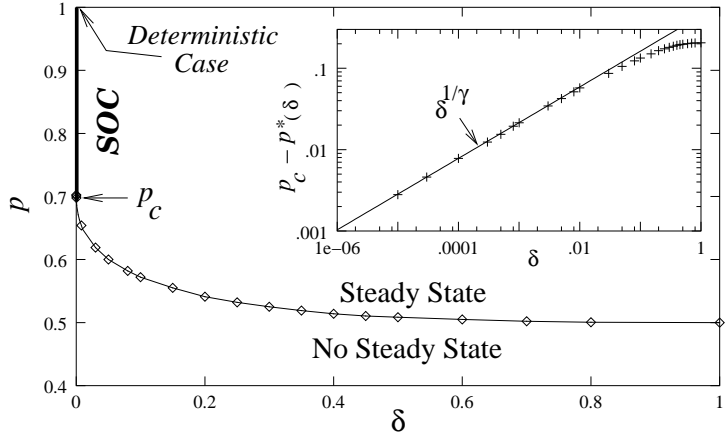}
\caption{The phase diagram in the $p$-$\d$ plane.  The inset shows the 
variation of $[p_c-p^*(\d)]$ versus $\d$, with the straight line showing  the 
theoretical fit (from [10]).}
\end{center}
\end{figure}

We then extended the arguments given to undirected sandpiles, by 
considering the time-evolution of a $d$-dimensional pile as a infection 
process on  a $(d +1)$-dimensional lattice. It can be proved that point 
i)-iv) listed above continue to hold for the undirected case, and   $ 
p^*(\delta =0)$   is exactly given by the critical threshold of the 
corresponding DP process. 

Again, if we can neglect correlations between heights of pile at different
sites in the steady state, the avalanche exponents in the model can be
seen to be same as DP exponents. Correlations between heights in this case
are stronger than in the directed case, as height at a site remains
unchanged if there is no activity, and is obviously correlated in time.
Thus, it is not clear that these are also irrelevant in the
renormalization group sense in our problem. In fact, Bonachela et al argue
that these are not. However, we note that because our toppling rules do
not depend on the height, so long as it is greater than $1$, most of these
correlations do not come into picture, as probabilities of different
topplings depend only whether the height at a site was zero or non-zero
before the addition of the new particle that made it unstable.  The
agreement of the results of our numerical simulations with the theoretical
predictions shows that indeed, for determining the critical exponents,
neglecting the effect of these correlations is justified.


Bonachela et al. have argued that while the arguments in \cite{mohanty_dhar} are
correct for the directed model, the neglect of correlations is not valid
in the case of undirected models. They presented Monte Carlo evidence
based on simulations of the fixed-energy sandpile (FES) model, and also
some nonrigorous arguments in support of their proposition.

Let us discuss the simulations of the FES first.  It should be noted that
one cannot get any avalanche exponents directly from such simulations, as
by definition, the FES shows a single avalanche that never stops, and
there is no distribution of avalanche sizes to quantify. The avalanche
exponents can only be inferred indirectly, by determining some other
exponents, and then using scaling theory to relate them to the
conventional avalanche exponents. For example, exponent governing the
decay of activity-activity correlation function with time in the steady
state of the fixed-energy sandpile is used to determine the exponent
$\beta/\nu_{\parallel}$.

The FES sandpile model can be obtained from our model by considering a 
uniform addition of particles at rate $\epsilon$ per site per unit time, 
with $\epsilon \ll \delta$,  and  take the 
double limit $ \epsilon \rightarrow 0$, and $\delta \rightarrow 0$. In the 
usual slowly driven sandpiles, one takes the limit $\epsilon \rightarrow  
0$ first, and then $\delta \rightarrow 0$ limit. In the FES case, we have 
to take the limit $\epsilon \rightarrow 0$ with mean activity $\bar{a} = 
\epsilon/(2\delta)$ held fixed, and then take the limit $\bar{a} \rightarrow 
0$. The different order of limits can lead to different scaling behaviors.

\begin{figure}
\begin{center}
\includegraphics[scale=.4]{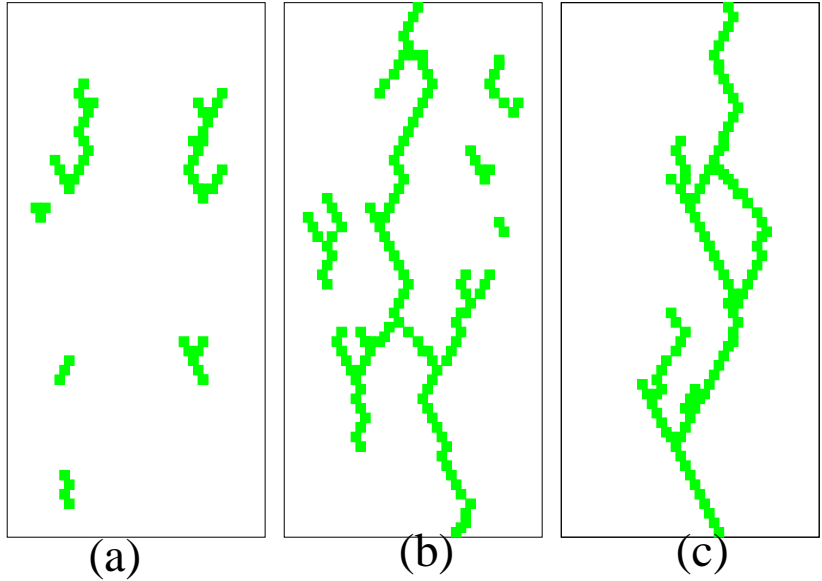}
\caption{ Schematic representation of time-evolution in the model in 
different cases: (a) non-overlapping avalanche clusters for very small   
$\epsilon  $ (b) an infinite  avalanche cluster and disconnected finite 
clusters for intermediate $\epsilon$, (c) Avalanche cluster in the fixed 
energy sandpile.}  
\end{center}
\end{figure}

In Fig. 2, we have shown a schematic space-time plot of the evolution of
the activity in a part of the sandpile in different cases (The y-axis is
time, and it increases upwards). In the case of slowly driven sandpile
case , with $\epsilon \ll \delta$, we see that different activity clusters
are disconnected from each other, and show a wide variation of sizes [
panel (a)]. If $\epsilon$ is comparable to $\delta$, there are overlapping
activity clusters [panel (b)]. However, there still are disconnected
clusters of activity, and one can find events where activity starts from
inert background ( due to addition of a particle), and the avalanche activity so
generated diffuses around, and may later die, or merge with the infinite
avalanche. In (c), we show the FES activity cluster. There are no
disconnected activity clusters, and the large-scale structure of the
cluster is different from (b), in that no activity can start from inert
state unless a neighbor was active. Thus, the large-scale structure of
clusters is clearly different in the three cases.

In particular, we note that the avalanche distributions in the
conventional slowly driven sandpile are determined by the properties of
disconnected finite clusters (when $\epsilon \ll \delta$, the probability
that the added new particle will be added to an active site is small). In
contrast, the FES properties are determined by the properties of the
single infinite cluster.

 We note, as argued in \cite{mohanty_dhar}, that on the line $p
=p^*(\delta)$, the probabilities of all clusters in the sticky sandpile
model are exactly those of DP clusters, with the infection probability
$\lambda$ given as an explicit function of $p$ and $\delta$.  This is so,
as on this line, the fraction of sites with height $0$ is zero in the
steady state, and every infected site has the same probability
$\lambda = p ( 1 - \delta)$ of infecting a downward neighbor.  If $p >
p^*(\delta)$ by a small amount $\Delta p$, the mean height would be large,
and concentration of sites with height $0$ in the steady state is small of
order $\Delta p$. 
The size of a cluster $s$ is defined in the space time plot as
the total  number of sites whose height has been altered during a
single  avalanche.
Clearly the difference in cluster size distribution  
$|{\rm Prob}(s) - {\rm Prob}_{DP}(s)|$ tend to
zero as $\Delta p$ tends to zero, and the probabilities for small
avalanches tend to the DP value near the boundary $ p = p^*(\delta)$.

 We  also note  the activity cluster in the FES for large average 
energy $E$ is  the same as the infinite DP-cluster with site 
percolation parameter $p$, as each site having at least one particle 
coming to it would topple with a probability $p$. Thus, in the
limit of large mean energy, all $n$-point correlation functions of the
activity in the steady state are exactly the same as in the DP process
with infection parameter $\lambda =p$.

 The only possibility for seeing the Manna exponents in the avalanche
statistics then  is to look for it in  some intermediate range of 
values of $s$ , $s_{min} \gg s \gg s_{max}$,  where $s_{min}$ is the 
crossover scale above which DP
behavior is presumably lost, and $s_{max}$ is the upper cutoff imposed by 
finite value of $\delta$. Here $s_{min}$ is expected to diverge as $
(\Delta p)^{-a}$ near the critical line. As the difference between the DP
and C-DP exponents is small, $s_{min}$ would be expected to be large, and 
the difference (even if present) is difficult to establish convincingly by
simulations.

 Bonachela et al. have also studied the problem using a numerical study of the
coupled Langevin equations for the density fields for activity and
particles. However, this study is also a numerical integration of the
stochastic evolution equations, and is qualitatively not different from a
Monte Carlo simulation working with coarse-grained variables instead of
the original variables defined on the lattice sites.

Finally, we discuss our more recent Monte Carlo data for the test case of
one-dimensional undirected sandpile model with sticky grains. We
monitored the average transverse cluster size as a function of the number
of topplings in the cluster. This is expected to be less sensitive to the
upper cutoff on the $s$. We did our simulations  for $p = 0.85, \delta =
10^{-5}$ on a line of length $L=4096$ with periodic boundary conditions.
This value of $\delta$ is a factor $10$ smaller than the values used in 
\cite{mohanty_dhar}, which allows us to generate much larger clusters.
The chosen value of $p$ (and a small $\delta$) is half-way between $p_c$ 
and $1$, and one would expect the effects of crossovers from DP and 
deterministic limits to be small.
The mean height in the steady state at this value of $p$ is $1.977(5)$, and
approximately $8.3\%$ of the sites are of zero height. Thus, any
deviations from the DP behavior coming from the presence of such sites
should be measurable. The results of data taken for $2\times 10^6$ avalanches are
shown in Fig. 3.  For comparison, we have also plotted the results for DP
clusters with $p = 0.7$, and for the Manna model. It is clear that the
observed slope is much closer to DP than to that for Manna clusters.

\begin{figure}
\begin{center}
\includegraphics[ width=4 in]{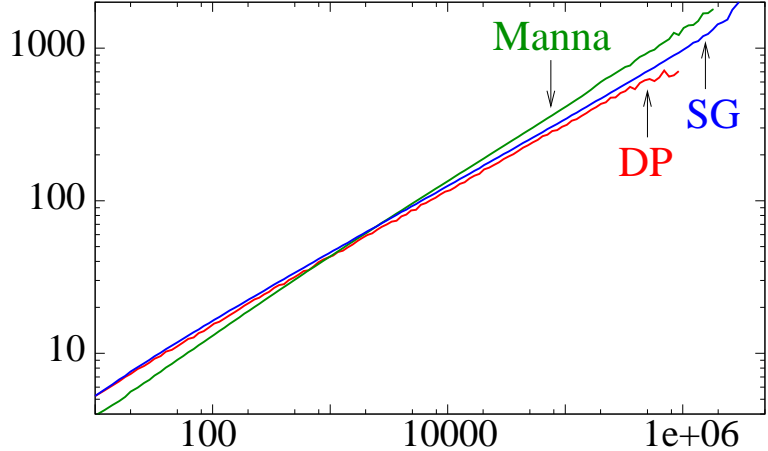}
\caption{ Log-log plot of mean transverse size of cluster   as a 
function of number of topplings in the cluster for the  sandpile model 
with sticky grains(SG) in 
one dimension for $p =0.85$ (blue line). For a comparison, we also plot 
simulation results for DP clusters (red line), and Manna model clusters 
(green line). }  
\end{center}
\end{figure}

   The theoretical arguments for neglecting, or not neglecting the
correlations are not fully convincing, and appeal to numerical simulations
for their justification. Extracting critical exponents from simulation
data could be complicated by crossover effects, as  the difference in the
avalanche exponents for the Manna and DP cases is not large.  We hope
that further work on this problem will clarify the situation, and lead to
better insight into the problem.




\begin{thebibliography}{99}
\bibitem{btw} P. Bak, C.Tang and K. Wiesenfeld, \prl {\bf 59}, 381 (1987);  
Phys. Rev. A {\bf 38}, 364(1988). 

\bibitem{soc} Some recent reviews are H. J. Jensen, {\it Self -Organised
Criticality} (Cambridge University Press, England, 1998);
E. V. Ivashkevich and V. B. Priezzhev, Physica A {\bf  254}, 97(1998);
F. Redig, Les Houches lectures, preprint, 2005; D. Dhar, Physica A {\bf 369}, 29
(2006).


\bibitem{granular} H. M. Jaeger and S. R. Nagel, Science {\bf 255}, 1523
(1992);{\it Granular Matter}, edited by A.  Mehta (Springer, Heidelberg,
1994); H. J. Herrmann, Physica A {\bf 263}, 51 (1999).


\bibitem{kadanoff} L.P.Kadanoff, S. R. Nagel, L. Wu and S. M. Zhou, Phys.
Rev. A {\bf 39} 6524 (1989).


\bibitem{manna1} S. S. Manna, Physica A {\bf  179}, 249(1991).


\bibitem{maslov} S. Maslov and Y-C. Zhang, Physica A {\bf 223}, 1 (1996).

\bibitem{vespignani} A. Chessa, E. Marinari and A. Vespignani, 
\prl {\bf 80}, 4217 (1998).

\bibitem{biham}{A. Ben-Hur and O. Biham, \pre {\bf 53}, R1317 (1996).}

\bibitem{brazil}  R. Dickman, M. A. Mun\~oz, A. Vespignani and
S. Zapperi, Brazilian J. Phys., {\bf 30}, 27 (2000).

\bibitem{mohanty_dhar} P. K. Mohanty and D. Dhar, Phys. Rev. Lett. {\bf 
89} (2002) 104303.

\bibitem{bonachela} J. A. Bonachela, J.  J. Ramasco, H. Chat\'e, I. 
Dornic, and M. A. Mun\~oz, Phys. Rev. E {\bf 74}, 050102 (2006).


\bibitem{dk}E. Domany and W. Kinzel, \prl {\bf 53}, 311 (1984).


\end{thebibliography}
\end{document}